\documentclass{PoS}
\usepackage{bm}
\newcommand{\beq}{\begin{equation}}
\newcommand{\eeq}{\end{equation}}
\newcommand{\bea}{\begin{eqnarray}}
\newcommand{\eea}{\end{eqnarray}}

\newcommand{\subrm}[1]{{\scriptscriptstyle\mathrm{#1}}}

\title{Lattice QCD Calculations with $\bm{b}$ Quarks:
Status and Prospects}
\ShortTitle{Lattice QCD Calculations with $b$ Quarks}

\author{\speaker{Matthew Wingate}\\
        Department of Applied Mathematics and Theoretical Physics, 
        University of Cambridge, Wilberforce Road, 
	Cambridge CB3 0WA, UK\\
        E-mail: \email{M.Wingate@damtp.cam.ac.uk}}

\abstract{This talk gives an overview of how lattice QCD calculations
are influencing quark flavor physics.
The first part of the talk focuses on the climb to higher precision; the
second part surveys views along less-trodden paths.}

\FullConference{The 13th International Conference on B-Physics at Hadron 
Machines (BEAUTY 2011)\\
April 4-8, 2011\\
Amsterdam, The Netherlands}

\begin{document}
\bibliographystyle{apsrev}

\section{Introduction}
\label{sec:intro}

It is a pleasure to give a brief review of lattice QCD (LQCD) calculations
relevant for quark flavor physics, particularly $b$ physics.  I am glad that
the other theoretical review talks were able to review results from sum rules,
models, and other nonperturbative approaches.  Effective field theory plays a
crucial role in LQCD, as in many other approaches; however, explicit
discussion here will be limited.  Even narrowing the scope of this talk to the
lattice, I am thankful to the organizers for the constraints in speaking time
and word count which preclude an encyclopedic review, thereby inviting an
idiosyncratic one.  In the following sections, I will say a few things about
how LQCD results are being used in tests of CKM unitarity and $B_s$ mixing,
and how I hope they can be used in rare $B_{(s)}$ and $\Lambda_b$ semileptonic
decays.

Many attendees at this conference are familiar with remarks introducing LQCD.
In the interest of brevity, allow me to just remind you that the following
hierarchy of scales
\begin{equation}
a ~\ll~ \left\{\frac{\hbar}{m_b c}, \frac{\hbar}{p_{\mathrm{hadron}}}, 
\frac{\hbar c}{\Lambda_{\subrm{QCD}}}, \frac{\hbar}{m_\pi c} \right\} ~\ll~ L
\end{equation}
is what is desired for the lattice spacing $a$ and box size $L$ in order to
have uncertainties of a purely statistical nature (leaving aside the
matching of regularization scheme-dependent quantities).  However,
what is achievable presently is closer to
\begin{equation}
\frac{\hbar}{m_b c} ~<~ a ~\ll~ \left\{ \frac{\hbar}{p_{\mathrm{hadron}}},
\frac{\hbar c}{\Lambda_{\subrm{QCD}}}\right\} ~\ll~ L ~<~ \frac{\hbar}{m_\pi c}
\end{equation}
with a restriction on $p_{\mathrm{hadron}}$, the spatial momentum of any
hadron in the lattice rest frame.  We numerically solve the physics of the
$\Lambda_{\subrm{QCD}}$ scale; then we apply HQET or NRQCD to treat the
physics of the $m_b$ scale, chiral perturbation theory to extrapolate to the
$m_\pi$ scale and estimate finite $L$ effects, and Symanzik effective theory
to reduce finite $a$ effects.  Operator matching between lattice and continuum
regularizations must often be done perturbatively, leading to another source
of truncation error.  The estimation and reduction of these uncertainties
occupies most of the LQCD effort in calculating matrix elements.

\section{CKM unitarity tests}
\label{sec:ckm}

I promised not to be encyclopedic, so I will not review individual lattice
calculations here.  Nevertheless, the list of recent unquenched calculations
of form factors for $|V_{ub}|$ \cite{Gulez:2006dt,Bailey:2008wp} and
$|V_{cb}|$ \cite{Okamoto:2004xg,Bernard:2008dn}; of the decay constants
$f_{B_{(s)}}$
\cite{Wingate:2003gm,Gamiz:2009ku,Blossier:2009gd,Albertus:2010nm,Simone:2010zz,Blossier:2010vj};
and of the $\Delta B = 2$ bag factors $B_{B_{(s)}}$
\cite{Dalgic:2006gp,Gamiz:2009ku,Albertus:2010nm} demonstrates the vast LQCD
effort being put into $B$ physics.  Similar effort is being dedicated to $K$
physics (\textit{e.g.}\ see \cite{Colangelo:2010et}), in particular the
$\Delta S = 2$ bag factor $B_K$
 \cite{Gamiz:2006sq,Aoki:2008ss,Antonio:2007pb,Aubin:2009jh,Aoki:2010pe,Bae:2010ki,Constantinou:2010qv},
 the $K\to \pi$ form factors \cite{Boyle:2007qe,Brommel:2007wn,Lubicz:2009ht,Boyle:2010bh},
 and the decay constants $f_\pi$ and $f_K$
 \cite{Beane:2006kx,Follana:2007uv,Aoki:2008sm,Aubin:2008ie,Bazavov:2009bb,Blossier:2009bx,Noaki:2010zz,Durr:2010hr}.

 The primary goal of these calculations at present is to feed into CKM fits,
 checking for a single allowed region in the 4-dimensional space of $(\lambda,
 A, \bar\rho, \bar\eta)$.  As we will see later, however, the same quantities
 enter other observables, which are not precisely determined enough to
 constrain the CKM fits, but which might turn out to differ from their
 Standard Model predictions.  Once the Standard Model has been supplanted, the
 hadronic matrix elements will still be needed in combination with whichever
 effective weak Hamiltonian replaces the one derived from the Standard Model.

 The plots displaying global fits to the CKM parameters $\bar\rho$ and
 $\bar\eta$ have become iconic.  Often we see them in the bold, outspoken
 colors used by CKMfitter. Other times they are displayed with the muted,
 conservative pastels used by the UTfit group.  New groups are joining in the
 artistry, choosing their own color schemes.  The talk by Lacker
 \cite{Lacker:2011xxxx} discusses these fits in detail, but I want to briefly
 emphasize some differences in how three groups, CKMfitter
 \cite{Tisserand:2009ja}, UTfit \cite{Lubicz:2008am,Bevan:2010zz}, and Laiho,
 Lunghi, Van de Water (LLV) \cite{Laiho:2009eu}, use lattice results in their
 fits.  (Sometimes a fit is done without LQCD data, which then allows a useful
 comparison between the fit results and lattice results.)

Before a quantity computed on the lattice can be taken as an input into the
CKM fit, decisions must be made about which LQCD results to average
together and how to propagate quoted uncertainties.  Naturally, different
groups tend to make different choices.

In the first respect, the difference between CKMfitter \cite{Tisserand:2009ja}
and UTfit \cite{Lubicz:2008am,Bevan:2010zz} on one hand and LLV
\cite{Laiho:2009eu} on the other is whether to include lattice results which
are missing a dynamical strange quark, \textit{i.e.}\ $N_f=2$ calculations
vs.\ the more physical $N_f=2+1$.  Empirically, it has not yet been shown that
the effects of the dynamical strange quark can be seen within present errors.
However, one cannot argue the effects are perturbatively small, whereas
one can for dynamical $c$, $b$, and $t$ quark effects.  The LLV group take the
latter view, not risking contamination from strange quark quenching effects.
The disadvantage is that, by excluding some calculations, some other lattice
systematics, \textit{e.g.}\ choice of discretization, are not averaged over to
the same extent.  Therefore, I find the CKMfitter/UTfit approach of including
$N_f=2$ results to be a reasonable compromise at the present time, although one
will want to adopt the LLV approach ultimately.  (In fact, for some quantities,
UTfit actually use the LLV averages.)

The second difference between groups is how they treat the systematic errors
assigned to individual lattice results.  This is not surprising since, for
example, some of these uncertainties arise due to the truncation of an
expansion in a small parameter (\textit{e.g.}\ the strong coupling $\alpha_s$
or $\Lambda_{\subrm{QCD}}/m_B$).  The size of the truncated terms is usually
estimated to be equal to the next higher power of the small parameter times a
number of $O(1)$.  How to propagate that estimation in a statistical analysis
is far from clear and groups make different choices.  UTfit describe some
lattice errors with a Gaussian distribution and others with a uniform
distribution \cite{Ciuchini:2000de}.  CKMfitter use their Rfit procedure
which treats the statistical error as a Gaussian distribution and the
systematic error as a flat distribution \cite{Hocker:2001xe,Tisserand:2009ja}.
The total error is propagated forward as a broadened, flattened Gaussian.  In
contrast, LLV treat the quoted systematic error as an independent Gaussian
distribution, to be combined in the usual way with the statistical error
\cite{Laiho:2009eu}.  The effective difference between these approaches is
that LLV usually quote more precise averages than CKMfitter.  Whether one
approach is too aggressive or the other is too conservative is a debate
unlikely to reach a conclusion.  In my experience, the dimensional analysis
method of estimating truncation errors is reliable; one sees $3\sigma$
discrepancies about as often as one would expect for a Gaussian distribution.
Therefore, I think the LLV fits produce fair error estimates.  Nevertheless it
is understandable that one would want to be a bit more cautious with this type
of theory error compared to a truly statistical error.

To give an idea how the different choices made by these groups can affect the
inputs to CKM fits, let us look at the $K^0$ mixing parameter $\hat{B}_K$;
respectively the groups quote $0.721(5)_{\mathrm{stat}}(40)_{\mathrm{sys}}$
(CKMfitter \cite{Tisserand:2009ja}), $0.725(27)$ (LLV \cite{Laiho:2009eu}),
and $0.731(7)_{\mathrm{stat}}(35)_{\mathrm{sys}}$ (UTfit
\cite{Lubicz:2010nx}).  The difference between a 5.5\% uncertainty (or larger)
and a 3.7\% uncertainty in $\hat{B}_K$ actually has interesting consequences.
With the smaller quoted uncertainty, a tantalizing discrepancy appears in CKM
fits to $\epsilon_K$, $\Delta M_s/\Delta M_d$, and the angle $\beta$
determined from $B\to (J/\psi)K_S$
\cite{Buras:2008nn,Lunghi:2008aa,Buras:2009pj,Laiho:2009eu}.  Also, with a 4\%
determination of $\hat{B}_K$, the uncertainty in $|V_{cb}|$ -- in particular
the exclusive/inclusive discrepancy -- becomes important.  This tension may be
due in part to NNLO corrections to $\epsilon_K$ which have now been
calculated \cite{Brod:2010mj}.

While much LQCD activity is going forward, verifying presently quoted
uncertainties, it is generally the case that new techniques are necessary to
have a discrete improvement in precision.  One area where we can expect
improvement soon is in $B \to D^{(*)}$ form factors.  Improved discretizations
appropriate for charm quarks \cite{Follana:2006rc,Oktay:2008ex} are now being
implemented.  As an indication of what can be achieved, the uncertainties in
$f_{D_{(s)}}$ have been decreased to the 1-2\% level
\cite{Follana:2007uv,Davies:2010ip}.  We can expect that this development,
along with the growing libraries of configurations being generated by the MILC
Collaboration \cite{Bazavov:2009bb} and others, will produce an improved
determination of $\mathcal{F}^{B\to D^{(*)}}$.  (Keep in mind that the
uncertainties coming from the $b$ quark on the lattice will degrade the
precision compared to $f_{D_{(s)}}$.)  See talks by Lacker and Mannel
\cite{Lacker:2011xxxx,Mannel:2011xxxx} which discuss CKM fits and their inputs
in more detail.

\section{$\bm{B_s}$ mixing}
\label{sec:mixing}

One hot topic, already reviewed at this conference
\cite{Bertram:2011xxxx} has been the measurement by the D0 experiment of an
anomalous like-sign di-muon asymmetry (due to one of the
$B_{(s)}^0\overline{B}^{\,0}_{(s)}$ pair oscillating before decaying) which is
50 times larger than expected from the Standard Model
\cite{Abazov:2010hv,Abazov:2010hj,Lenz:2006hd,Lenz:2011ti}, a discrepancy at
the $3\sigma$ level:
\begin{eqnarray}
A^{\mathrm{D0}}_{SL} &=& (-9.6 \pm 2.5 \pm 1.5)\times 10^{-3} \\
A^{\mathrm{LN}}_{SL} &=& (-0.20 \pm 0.03)\times 10^{-3}
\end{eqnarray}
where the asymmetry is roughly an average of flavor-specific asymmetries
$A_{SL} \approx (a^d_{\mathrm{fs}} + a^s_{\mathrm{fs}})/2$. Given the size of
the experimental signal and uncertainty, this is still something for the
experimentalists to pin down.  From my perspective though, it is interesting
to note that the leading uncertainty in the theoretical estimate comes from
the uncertainty in $|V_{ub}/V_{cb}|$ \cite{Lenz:2006hd,Lenz:2011ti}.  In
2006/07 this uncertainty contributed to a 20\% uncertainty in the
flavor-specific asymmetry $a^s_{\mathrm{fs}}$.  By 2010/11 the
$|V_{ub}/V_{cb}|$ error only contributed to a 12\% uncertainty in
$a^s_{\mathrm{fs}}$.  The improvement in part came from progress combining
much-improved theoretical and experimental work on $B\to \pi\ell\nu$ (see
\cite{Mannel:2011xxxx} here).

Turning to measurements which are limited by theoretical uncertainties, let me
focus on Standard Model calculations of $\Delta M_s$ and $\Delta \Gamma_s$.
The $B_s$ decay constant $f_{B_s}$ is the most important hadronic quantity
entering the mass and width differences; it enters because one writes the
matrix element of the 4-quark operators as the product of $f_{B_s}^2$ times
known factors, which would be the result in the vacuum saturation
approximation (VSA), and the ``bag-factors'' which represent deviations from
the VSA.  According to Lenz and Nierste's analysis, the contribution of the
$f_{B_s}$ uncertainty to the errors in $\Delta M_s$ and $\Delta \Gamma_s$ has
decreased from 33\% to 13\% in the past 5 years due to progress in lattice QCD
\cite{Lenz:2006hd,Lenz:2011ti}. There is still much room for improvement;
$f_{B_s}$ is still the most uncertain quantity in the calculation of $\Delta
M_s$; the uncertainty in $B_{B_s}$ is even slightly less important than the
$|V_{cb}|$ uncertainty, although the latter is more likely to be see a
significant reduction soon.  In the case of the width difference, $f_{B_s}$ is
now precise enough that further reduction of uncertainty in $\Delta \Gamma_s$
requires a full lattice calculation of matrix elements through
$O(\Lambda_{\subrm{QCD}}/m_b)$.  By itself, the numerical computation of these
matrix elements would not be difficult; it is the perturbative matching
calculation which requires significant, dedicated human effort.

\section{Rare decays}
\label{sec:rare}

We now turn from rare mixing to rare decays, already reviewed at this
conference \cite{Straub:2011xxxx}.  While the search for new physics in CKM
fits is in some ways like a difficult, technical climb to the summit of Mt.\
Precision, the search for new physics in rare decays is like bushwhacking into
the wilderness of overgrown backgrounds and nonfactorizable snakes.  With
such stealthy prey, we need hunters covering all paths.

The rare decays $B \to K^* \gamma$ and $B\to K^{(*)}\ell^+ \ell^-$ have been
measured at the Tevatron and $B$ factories \cite{Asner:2010qj} and their
branching fractions agree with Standard Model estimates
\cite{Ball:2006eu,Ali:1999mm}.  
CDF has recently observed $B_s \to \phi \mu^+\mu^-$ \cite{Aaltonen:2011cn}.
The LHC experiments (especially LHCb) expect
to more precisely measure the lepton invariant-mass spectrum of these decays
as well as other observables, some of which may reveal signs of physics beyond
the Standard Model.  While some of these observables are constructed so that
hadronic quantities cancel, others will need precise determinations of
hadronic matrix elements \cite{Bobeth:2010wg,Bobeth:2011gi}.

Here lattice calculations can help by computing the form factors which
parametrize the various hadronic matrix elements.  Essentials of the
calculations are the same as in the calculation of $\langle \pi(p') | V_\mu |
B(p)\rangle$ \cite{Gulez:2006dt}, except a new matching calculation is
necessary for the tensor operator \cite{Muller:2010kb}.

The theory of rare $B_{(s)}$ decays at large recoil is under good control
\cite{Buchalla:1998mt,Grinstein:2004vb,Beylich:2011aq}.  The main cause for
concern has been nonlocal effects, primarily arising from the operator $Q_2 =
(\bar{s} b)_{V-A}(\bar{c}c)_{V-A}$, which creates a charmonium resonance before
decaying to a lepton pair.  At sufficiently large $q^2$ the matrix elements of
non-local operators can be written in terms of the form factors in an operator
product expansion.  Rare, exclusive $b\to s\ell^+\ell^-$ decays
at low recoil look to be a promising new place to test the Standard Model.

This kinematic range is exactly where LQCD calculations can be done.  As $q^2$
decreases, we first encounter growing discretization errors as the spatial
momentum $|\mathbf{p}'|$ of the final state meson becomes comparable to the
inverse lattice spacing.  While some tricks can be played
\cite{Horgan:2009ti}, HQET errors also grow like $v\cdot p'/m_B$ (where $v$ is
the 4-velocity of the $B$).  My collaborators and I have performed a
calculation of the 10 form factors governing $B_{(s)}$ semileptonic and rare
decays, including $SU(3)_F$ breaking effects. Preliminary results have
appeared in conference proceedings, most recently CKM2010 \cite{Liu:2011ra}.
Very soon we should be finalizing our calculations.

\section{Beautiful baryons}
\label{sec:baryon}

The higher energy of LHC collisions will allow us to become better acquainted
with the properties of baryons which contain a $b$ quark, building upon the
discoveries of the Tevatron experiments \cite{Asner:2010qj}.  At the same 
the first experimental observations of many of these states were being made,
the masses of the $b$-baryons were computed in unquenched lattice QCD 
\cite{Na:2007pv,Lewis:2008fu,Burch:2008qx,Na:2008hz,Detmold:2008ww,Lin:2009rx,Wagner:2011fs}.
A compilation of these results show good agreement among the computations,
which use a variety of lattice formulations \cite{Lewis:2010xj}.

The study of $\Lambda_b \to \Lambda \ell^+\ell^-$ at the LHC is also a
promising one.  The short distance physics should be the same as in $B \to
K^{(*)}\ell^+\ell^-$.  Nevertheless, with the $\Lambda$ in the final state,
one hopes the $\Lambda$ baryon's polarization will increase sensitivity to any
new right-handed couplings \cite{Chen:2001ki,Aliev:2002ww}. In LQCD one finds
worse signal-to-noise ratios in correlation functions involving baryons
compared to pseudoscalar mesons, so it would be difficult to extract all 12
form factors governing $\Lambda_b \to \Lambda$ decays.  However, in the heavy
quark limit these 12 reduce to only 2 \cite{Mannel:1990vg}:
\begin{equation}
\langle \Lambda(p')| \bar{s}\;\Gamma\, b | \Lambda_b(p)\rangle ~=~
\bar{u}_\Lambda(p')\Big[F_1(q^2) \;+\; {\rlap{/}\kern-1.0pt v}F_2(q^2)\Big] 
\Gamma \, u_{\Lambda_b}(p)
\end{equation}
where $q = p-p'$.  The time is ripe for LQCD study of matrix elements
of this type.

\section{Conclusions}
\label{sec:concl}

We have every reason to believe there are natural explanations for the
peculiarities of the Standard Model.  It is clear we need as much information
as possible -- experimental data and theoretical calculations -- in order to
find overt signs of new physics in some places and to be sure of its absence
in other places.  We must pursue every path and leave no stone unturned.

\section*{Acknowledgments}

I am grateful for support from the Science \& Technology Facilities Council
and the Institute for Particle Physics Phenomenology, Durham University.



\bibliography{mbw}

\end{document}